\documentstyle[aps]{revtex}
\input{psfig}
\draft

\begin{document}
\wideabs{
\title{Josephson effect and tunneling spectroscopy in Nb/Al$_{2}$O$_{3}$/Al/MgB$_{2}
$ thin films junctions}
\author{G. Carapella$^{1}$, N. Martucciello$^{1}$, G. Costabile$^{1}$, C. Ferdeghini$%
^{2}$, V. Ferrando$^{2}$, and G. Grassano$^{2}$}
\address{
$^{1}$INFM Research Unit and Department of Physics ''E. R. Caianiello'',
University of Salerno, I-84081 Baronissi, Italy.}
\address{
$^{2}$INFM Research Unit and Department of Physics, University of Genova,
I-16146 Genova, Italy.}
\date{\today}
\maketitle

\begin{abstract}
We report the demonstration of dc and ac Josephson effects as well as
tunneling spectroscopy measurements on Nb/Al$_{2}$O$_{3}$/Al/MgB$_{2}$ thin
films heterostructures. Data on dc Josephson effect suggest the presence of 
two characteristic critical currents possibly accounted for the presence of
two gaps in the MgB$_{2}$ film. Tunneling spectroscopy measurements confirm
the two-gap scenario and are explained with the presence of tunneling both
from dirty limit regions, reflecting an order parameter $\Delta
_{dirty}\approx 2$~meV, and from clean limit regions, reflecting an order
parameter $\Delta _{3D}\approx $0.9~meV. The temperature dependence, the
magnitudes and the relevant ratios of these two gaps are found in good
agreement with the predictions of a recently proposed multigap theory.  The
value of the larger gap component estimated from  conductance spectra is
found to be consistent with the one deduced from  Josephson coupling.
\end{abstract}

}

The recent demonstration of superconductivity \cite{Nagamatsu} at 39 K in MgB%
$_{2}$ has attracted considerable interest with regard to both fundamental
issues and practical applications. Though several quasiparticle tunneling
experiments \cite
{Karapetrov,Schmidt,Rubio,Kohen,Plecenik,Szabo,Sharoni,Giubileo1,Giubileo2,Gonnelli2}%
with normal contacts have been performed to measure the energy gap, only few
experiments \cite{Gonnelli1,Zhang,Brinkman,Burnell} to probe the Josephson
effects in this new material have been performed. To date, only nanobridges 
\cite{Brinkman} and localized ion damage Josephson junctions \cite{Burnell}
have been successfully tested on MgB$_{2}$ films. To fully exploit the
possibility of superconducting electronic to cryocooler temperature promised
by this new material, a further effort toward realization and study of
Josephson junctions based on MgB$_{2}$ films is again necessary.

In this letter, we report Josephson effect and tunneling spectroscopy
analysis on Nb/Al$_{2}$O$_{3}$/Al/MgB$_{2}$ thin films junctions. At our
knowledge, this is the first demonstration of Josephson effects in a MgB$%
_{2} $ film based heterostructure. Experimental data on dc Josephson effect
indicate the presence of two classes of hysteretic Josephson junctions with
two characteristic critical currents accounting for probable presence of two
gaps in the MgB$_{2}$ film. Experimental data on tunneling spectroscopy
measurements confirm this two-gap scenario and are accounted for the
presence of tunneling both from dirty limit regions, reflecting an order
parameter $\Delta _{dirty}$, and from clean limit regions, reflecting an
order parameter $\Delta _{3D}$. The temperature dependence deduced from
tunneling spectroscopy gives for the dirty limit region $\Delta
_{dirty}(0)\approx 2$ meV, $T_{c}^{dirty}=14.5$ K and for the clean limit
region $\Delta _{3D}(0)\approx 0.9$ meV, $T_{c}\approx 29.5$ K , with
measured ratios $2\Delta _{dirty}(0)/k_{B}T_{c}^{dirty}=3.5,$ $\Delta
_{dirty}/\Delta _{3D}\approx 2.2$ in good agreement with the predictions of 
the recently proposed multigap theory \cite{Liu} of Liu {\it et al}.

{\em The junctions- }The MgB$_{2}$ films were deposited at room temperature
on Al$_{2}$O$_{3}$ substrate by Pulsed Laser Deposition (PLD), starting from
stoichiometric target. We used sintered target prepared by direct synthesis
from the elements. The PLD experimental apparatus consists of an UHV
deposition chamber and a KrF excimer laser; details of the apparatus are
described elsewhere \cite{Cimberle}. The as-deposited amorphous films, about
500 nm\ thick, were ex-situ annealed in magnesium vapor. The samples were
placed in a sealed tantalum tube, in Ar atmosphere, containing Mg lumps
(approximately 0.05 mg/cm$^{3}$), and then they were placed in an evacuated
quartz tube and heated at $T=$1120 K for 30 minutes followed by rapid
quenching at room temperature. The MgB$_{2}$ films resulted to be strongly
c-axis oriented as determined using synchrotron radiation at the ID32
beamline of ESRF. The films show a surface RMS roughness of about 15 nm\ as
proved by AFM measurements, a critical temperature ranging from 33 K to 37 K
and a maximum residual resistivity ratio of 2.5. Further details on the
electrical, structural and morphological film characterization are reported
in Ref. \cite{Ferdeghini}.

\begin{figure}[tbp]
\psfig{file=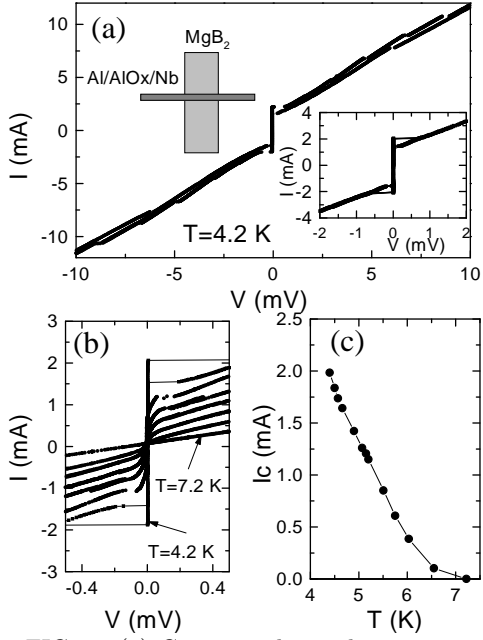,width=6.5cm,clip=}
\caption{(a) Current-voltage characteristic of the junction recorded at
$T=$4.2~K showing resistive branches starting with different Josephson
critical currents. In the insets, the IV curve with larger critical current
and a sketch of the junction geometry are shown. (b) IV curves recorded at
temperatures increasing from $T=$4.2~K to $T=$7.2~K. (c) Josephson critical
current versus temperature.}
\label{fig1}
\end{figure}

To realize the junctions, we avoided wet photolithographic techniques, that
could damage the surface of the MgB$_{2}$. For the bottom electrode (MgB$_{2}
$) definition, we developed a special setup inside the growth chamber for
the in-situ use of tantalum shadow masks. Therefore the precursor was
directly deposited in shape of narrow strip (about 200 $\mu $m wide) and
then annealed in Mg atmosphere. Before the definition of the superconducting
counter-electrode (Nb), the MgB$_{2}$ film was sputter-etched for about 10
nm in Ar atmosphere, to reduce the thickness of the weakened surface layer
growth after exposition to air. An Al/Al$_{2}$O$_{3}$/Nb structure was then
fabricated in-situ on the whole substrate. The 10~nm thick Al and the 500 nm
thick Nb films were rf sputtered while the $\sim $2~nm thick Al$_{2}$O$_{3}$
was thermally growth in pure oxygen atmosphere. The Al/Al$_{2}$O$_{3}$/Nb
structure was finally patterned by Reactive Ion Etching with the geometry of
the counter-electrode, a 40~$\mu $m wide strip. The cross geometry [see
inset of Fig.~\ref{fig1}(a)] of the fabricated junctions allows us to make
four contacts measurements. We fabricated and tested three of these
junctions with similar results. The heterostructures showed a good thermal
cyclability. For the junction we report here the critical temperature of the
Nb counter-electrode was $T_{c}^{Nb}$ =7.2~K and the critical temperature of
the MgB$_{2}$ electrode was $T_{c}^{MgB_{2}}$=33~K.

{\it Josephson effect}- Figure 1(a) shows the current-voltage (IV) curve of
the junction recorded at $T$=4.2~K. The current branch a $V=0$ clearly
suggests the evidence for a dc Josephson effect in the junction. However,
the resistive characteristic appears quite complex. Two distinct resistive
families can be observed, starting with two slightly different critical
Josephson currents. Within the same resistive family, slightly different
resistive branches are exhibited, meaning that we are in fact concerned with
a disordered array of smaller Josephson junctions with comparable critical
currents. Moreover, also the magnetic field diffraction pattern we observed
is typical of a disordered array of small area Josephson junctions: very
irregular, with fast field modulation over a quite large current background.

Curves in Fig.~\ref{fig1}(a) are recorded with a 100~Hz current sweep, so
that a superposition of both families is plotted. In dc biasing, it is
possible to record one of the two resistive families, mainly the curve
corresponding to the larger Josephson critical current, as shown in the
inset of Fig.~\ref{fig1}(a). The evolution in temperature of such hysteretic
IV curve is shown in Fig.~\ref{fig1}(b). The quite large thermal smearing
envisaged in the IV curves at 4.2~K is compatible with the idea that our
junction consists of a parallel arrays of about 200 small junctions with
individual critical currents of about 10 $\mu $A. The total critical current
as a function of the temperature is plotted in Fig.~\ref{fig1}(c). The
behavior is qualitatively similar to the one exhibited \cite{Barone} in
proximity coupled $S_{1}INS_{2}$ structures, with a $N$ layer thicker than
the 10~nm Al layer we deposited. This means that the normal surface layer on
the MgB$_{2}$ was not completely removed with our sputter etch cleaning
procedure.

Figure \ref{fig1}(a) seems indicate the presence of two qualitatively
different Josephson junctions classes, possibly corresponding to two energy
gaps in the MgB$_{2}$ film. In this framework, an estimate of the larger gap
can be made from IV curve shown in the inset of Fig.~\ref{fig1}(a),
corresponding to the larger of the two possible critical currents we can
envisage in the main panel. Here, by assuming comparable values for the
critical currents of the junctions in the array , we can estimate a mean
characteristic voltage $V_{c}\equiv IcR_{N}$ $\approx 2.3$ mV. With a
critical temperature of 7.2 K, a gap $\Delta _{1}\simeq 1.1$ meV is found
for our Nb. The gap component $\Delta _{2}$ of MgB$_{2}$ responsible of the
higher critical current Josephson junction class can then estimated from
relation \cite{Barone} 
\begin{equation}
I_{c}R_{N}\simeq \frac{\pi }{e}\frac{\Delta _{1}\Delta _{2}}{\Delta
_{1}+\Delta _{2}}  \label{IC}
\end{equation}
as $\Delta _{2}=(2.1\pm 0.1$) meV at 4.2 K.

\begin{figure}[tbp]
\psfig{file=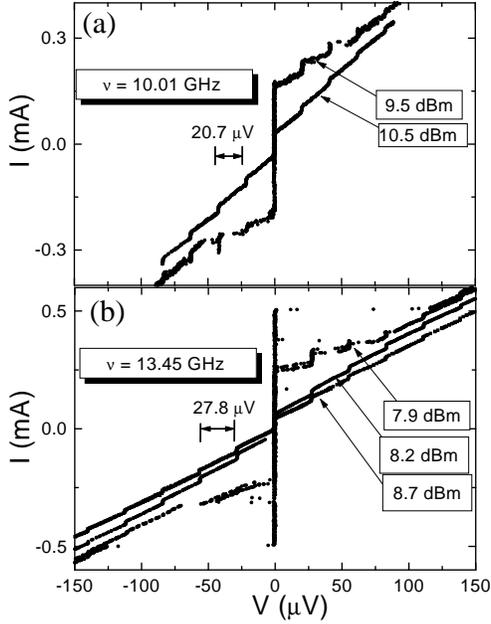,width=6.5cm,clip=}
\caption{IV curves of the junction showing Shapiro steps induced by a
microwave signal with different power level. The frequency is $\nu $%
=10.01~GHz in (a) and $\nu $=13.45~GHz in (b). The curves are recorded at
$T=$4.2~K.}
\label{fig2}
\end{figure}

The Josephson nature of the $V=0$ branch exhibited in our IV curve is
confirmed by the regular appearance of Shapiro steps (ac Josephson effect)
when the junctions is irradiated with rf signals in the microwave range. The
steps are expected at voltages V$_{n}=n\Phi _{0}\nu $, where $n$ is an
integer, $\Phi _{0}$ is the flux quantum, and $\nu $ is the frequency of the
applied radiation. Figure~\ref{fig2} shows these steps induced in the IV
curve of the junction irradiated with microwaves signals of different power
levels. As expected, the voltage spacing of the steps is found to be 20.7~$%
\mu $V for microwave power at frequency $\nu =10.01$~ GHz [Fig.~\ref{fig2}%
(a)], and 27.8~$\mu $V at frequency $\nu =13.45$~GHz [Fig.~\ref{fig2}(b)].

\begin{figure}[tbp]
\psfig{file=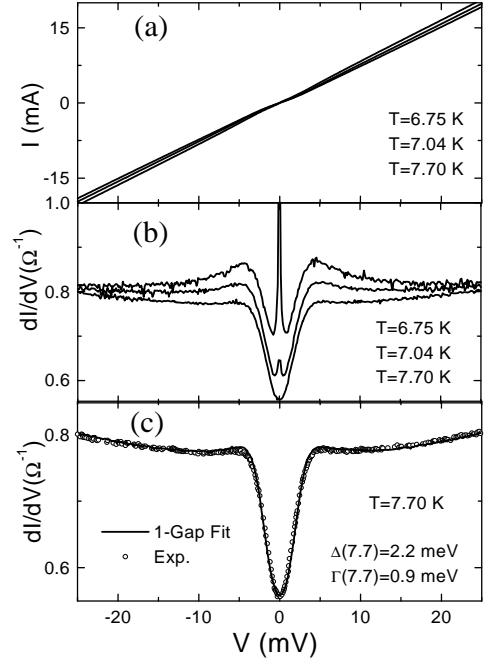,width=6.5cm,clip=}
\caption{ IV curves (a) and differential conductances (b) at temperatures
around the critical temperature of the Nb electrode. (c) Best fit (solid
line) of the differential conductance curve (open circles) at $T=$7.7~K with
the one-gap model.}
\label{fig3}
\end{figure}

{\it Tunneling spectroscopy}- Figure~\ref{fig3}(a) shows the IV curves
recorded at temperatures near the transition temperature of the Nb
electrode. The curves are now quite smooth, as better seen from differential
conductances $dI/dV$ shown in Fig.~\ref{fig3}(b). These conductance curves
were obtained by numerically differentiating the curves in Fig.~\ref{fig3}%
(a). When the Josephson peak at $V=0$ completely disappears, we are
concerned with a $N_{1}IS_{2}$ junction ($N_{1}=$ Nb in the normal state, $%
S_{2}=$ MgB$_{2}$ in the superconducting state) and a tunneling spectroscopy
of the proximized surface layer of the MgB$_{2}$ is possible.

The differential conductance $G(V)\equiv dI/dV(V$) can be fitted using the
functional form for a $NIS$ contact 
\begin{equation}
G \sim \frac{d}{dV}\int_{-\infty }^{\infty }\rho (E+eV)\left[
f(E)-f(E+eV)\right] dE,  \label{conduct}
\end{equation}
where $\rho (E)=Re \left\{ \left( E-i\Gamma \right) / \left[ \left(
E-i\Gamma \right) ^{2}-\Delta ^{2}\right] ^{1/2}\right\}$ is the modified 
\cite{Dynes} BCS density of states, $\Delta (T)$ is the fitted energy gap, $%
\Gamma (T)$ is the fitted gap-smearing parameter, and $f(E)$ is the Fermi
function. The best one-gap fit of the conductance curve of our junction a $T$%
=7.7 K is shown in Fig.~\ref{fig3}(c). For this temperature $\Delta =(2.2$ $%
\pm 0.2)$ meV, $\Gamma =0.9$ meV are estimated for the MgB$_{2}$ film.

Figure 4(a) shows the normalized differential conductances curves for
temperature increasing up the critical temperature of the MgB$_{2}$
electrode. The small parabolic background envisaged in the curves suggests a
moderate barrier height for the contact, probably below 100~meV. Moreover,
the quite large zero bias conductance means the presence of appreciable
leakage currents. The gap structure completely disappears at $T$=29.5 K,
lower than the critical temperature of 33~K measured for the bulk electrode,
meaning that we are probing proximized surface layer of the MgB$_{2}$ with
weakened superconductivity.

\begin{figure}[tbp]
\psfig{file=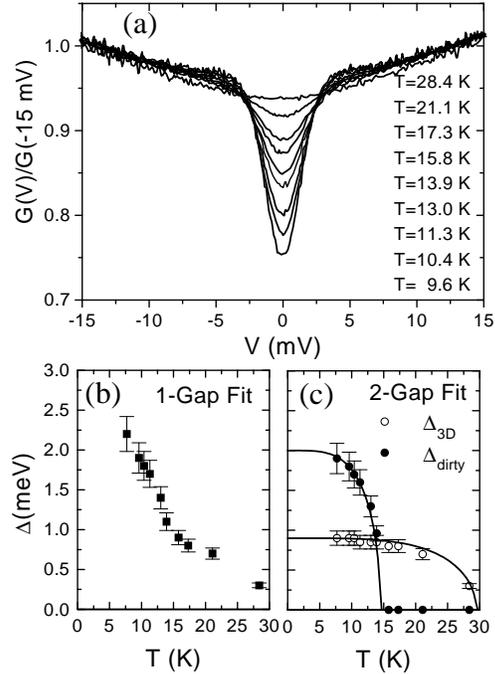,width=6.5cm,clip=}
\caption{(a) Normalized differential conductance curves for increasing
(bottom to top) temperatures. (b) Temperature dependence of the order
parameter from fit with the one-gap model. (c) $\Delta _{dirty}(T)$ and $%
\Delta _{3D}(T)$ from fit with the two-gap model. Fit to BCS curves (solid
lines) is also shown .}
\label{fig4}
\end{figure}

From the fit of the conductance curves in Fig.~\ref{fig4}(a) with one-gap
model Eq.~(\ref{conduct}) the temperature dependence of the energy gap shown
in Fig.~\ref{fig4}(b) is obtained. The curves are reasonably well fitted
using a gap-smearing parameter $\Gamma (T)\simeq 0.4\Delta (T).$ The
obtained temperature dependence of the energy gap clearly deviates from
single gap BCS behavior. Rather, data in Fig.~\ref{fig4}(a) suggest the
superposition of two BCS-like gaps with two different critical temperatures,
a behavior very similar to the one recently observed \cite{Plecenik} in
tunneling spectroscopy on MgB$_{2}$ wires. Indeed, many recent experimental
results from tunneling measurements \cite{Plecenik,Szabo,Giubileo1,Giubileo2}%
, high resolution photoemission spectroscopy \cite{Tsuda}, microwave surface
resistance \cite{Zhukov}, and heat specific measurements \cite{Wang}, are
explained within a two-gap scenario in MgB$_{2}$. On the other hand, many
other tunneling and point-contact spectroscopy measurements \cite
{Karapetrov,Schmidt,Rubio,Kohen,Sharoni,Gonnelli2,Zhang} are explained
assuming a single BCS-like gap in MgB$_{2}.$

The multigap model proposed by Liu {\em et al.} \cite{Liu} could explain
this complex experimental scenario. In the clean limit \cite{Liu}, two
different order parameters exist, a gap $\Delta _{2D}$ accounting for
current transport along B planes (a-b planes), and a gap $\Delta _{3D} $
accounting for current  transport in direction perpendicular to the B layers
(c-axis). The gap $\Delta _{3D} $ is three times smaller than the gap $%
\Delta _{2D} $ and both gaps close to the critical temperature $T_{c}$ of
the MgB$_{2}, $ following a BCS-like temperature dependence. Moreover, the $%
2\Delta _{2D}(0)/k_{B}T_{c}$ ratio is expected larger than the BCS value,
while the $2\Delta _{3D}(0)/k_{B}T_{c}$ ratio is expected lower than this
value. In the dirty limit \cite{Liu}, the enhanced defect scattering leads
to an averaging of both order parameters that results to an isotropic order
parameter $\Delta _{dirty}$ closing to a critical temperature$T_{c}^{dirty}$
lower than the clean limit $T_{c}.$ The magnitude $\Delta _{dirty}$ is
between the magnitudes of $\Delta _{2D}$ and $\Delta _{3D},$ and the $%
2\Delta _{dirty}(0)/k_{B}T_{c}^{dirty}$ ratio is the BCS value.

As stated above, our MgB$_{2}$ films resulted to be strongly c-axis
oriented, and from geometry of our junction a tunneling along c-axis is
expected, with spectra reflecting symmetry of $\Delta _{3D}$ order
parameter. On the other hand, due to the quite large area ($200\times 40$) $%
\mu $m$^{2}$ we are probing, we cannot exclude the presence of dirty regions
in the contact area. The presence of these dirty regions is compatible with
the quite large values of smearing parameter $\Gamma $ we have found. Thus
our junction can be considered as the parallel connection of two junctions
corresponding to the tunneling into the dirty and clean regions,
respectively. Thus, our measured conductance can be described by 
\begin{equation}
G(V)=\alpha G(V,\Delta _{3D})+(1-\alpha )G(V,\Delta _{dirty}),
\label{twogap}
\end{equation}
where $0<\alpha <1$, and the functional form $G(V)$ is again Eq.~(\ref
{conduct}). The temperature dependence of the two gaps deduced from the fit
of the data in Fig.~\ref{fig4}(a) with the the two-gap model Eq.~(\ref
{twogap}) is shown in Fig.~\ref{fig4}(c). The best fit was obtained with $%
\alpha =0.5$, meaning that clean limit and dirty limit regions was present
in equal proportions in our tunnel contact. The fit with BCS functional
temperature dependence gives for the clean limit $\Delta _{3D}(0)=(0.9\pm
0.1)$ meV, $T_{c}\approx 29.5$ K and for the dirty limit $\Delta
_{dirty}(0)=(2.0\pm 0.2)$ meV, $T_{c}^{dirty}\approx 14.5$ K, with measured
ratios $2\Delta _{dirty}(0)/k_{B}T_{c}^{dirty}\simeq 3.5,$ $\Delta
_{dirty}/\Delta _{3D}\approx 2.2$ in good agreement with the predictions of
Liu {\em et al}.

The two gaps we have invoked for the explanation of two Josephson critical
currents can be now identified with $\Delta _{dirty}$ and $\Delta _{3D},$
with $\Delta _{dirty}$ accounting for the larger Josephson critical current.
In fact, from $I_{c}R_{N}$ product we have estimated the larger gap
component amplitude $\Delta _{2}$ $=(2.1\pm 0.1$) meV, that is consistent
with the $\Delta _{dirty}(0)=(2.0\pm 0.2)$ meV we have found from tunneling
spectra analysis.

Summarizing, we have demonstrated dc and ac Josephson effect and we have
analyzed tunneling spectroscopy measurements in a rather large area Nb/Al$%
_{2}$O$_{3}$/Al/MgB$_{2}$ thin films junction we have fabricated. The
analysis of the dc IV curves has indicated the presence two characteristic
Josephson critical currents, probably accounted for the presence of two gaps
in the MgB$_{2}$ film. The tunneling spectroscopy measurements have
confirmed this two-gap scenario and have been explained in the framework of
the multigap theory of Liu {\em et al} with the presence of tunneling both
from dirty limit regions, reflecting an order parameter $\Delta
_{dirty}\approx 2$ meV, and from clean limit regions, reflecting an order
parameter $\Delta _{3D}\approx 0.9$ meV. Experimental data have been found
in good quantitative agreement with the predictions of Liu {\em et al}.

We acknowledge Dr. F. Giubileo and Dr. F. Bobba for useful discussions on
tunneling spectra analysis. Financial support of MURST COFIN00 is also
acknowledged.



\end{document}